\documentclass[conference]{IEEEtran}
\IEEEoverridecommandlockouts
\usepackage{verbatim}
\usepackage{longtable}
\usepackage{adjustbox}
\usepackage{listings}
\usepackage{xcolor}

\lstdefinelanguage{Solidity}{
  morekeywords={pragma, solidity, contract, struct, event, function, mapping, address, bytes, bytes32, bool, external, public, require, emit, calldata, payable, return, returns, if, else, memory, keccak256},
  sensitive=true,
  morecomment=[l]{//},
  morecomment=[s]{/*}{*/},
  morestring=[b]"
}

\usepackage{amsmath}
\usepackage{amsfonts}
\usepackage{amssymb}
\usepackage{multirow}
\usepackage{amsmath,amssymb,amsfonts}
\usepackage{fancyhdr}

\usepackage{graphicx}
\usepackage{textcomp}
\usepackage{xcolor}
\usepackage{enumitem}
\usepackage{float}
\usepackage{graphicx}
\usepackage[normalem]{ulem}
\usepackage{url}
\useunder{\uline}{\ul}{}
\usepackage{cite}
\usepackage{amsmath,amssymb,amsfonts}
\usepackage{graphicx}
\usepackage{textcomp}
\usepackage{comment}
\usepackage{subcaption}

\usepackage{xcolor}
\def\BibTeX{{\rm B\kern-.05em{\sc i\kern-.025em b}\kern-.08em
    T\kern-.1667em\lower.7ex\hbox{E}\kern-.125emX}}
\usepackage{algorithm}
\usepackage[noend]{algpseudocode}

\begin{document}

\title{\textit{CrossLink}: A Decentralized Framework for Secure Cross-Chain Smart Contract Execution}


\author{\IEEEauthorblockN{Tahrim Hossain$^{1}$, Faisal Haque Bappy$^{2}$, Tarannum Shaila Zaman$^{3}$, and Tariqul Islam$^{4}$}
\IEEEauthorblockA{
$^{1, 2, 4}$ Syracuse University, NY, USA, and
$^{3}$ University of Maryland, Baltimore County, MD, USA
\\
Email: mhossa22@syr.edu, fbappy@syr.edu, zamant@umbc.edu, and mtislam@syr.edu} 
}

\maketitle

\thispagestyle{fancy}
\lhead{\small{This work has been accepted at the 2025 IEEE ICBC Cross-Chain Workshop (ICBC-CCW 2025)}} 
\cfoot{}

\begin{abstract}
This paper introduces \textit{CrossLink}, a decentralized framework for secure cross-chain smart contract execution that effectively addresses the inherent limitations of contemporary solutions, which primarily focus on asset transfers and rely on potentially vulnerable centralized intermediaries. Recognizing the escalating demand for seamless interoperability among decentralized applications, \textit{CrossLink} provides a trustless mechanism for smart contracts across disparate blockchain networks to communicate and interact. At its core, \textit{CrossLink} utilizes a compact chain for selectively storing authorized contract states and employs a secure inter-chain messaging mechanism to ensure atomic execution and data consistency. By implementing a deposit/collateral fee system and efficient state synchronization, \textit{CrossLink} enhances security and mitigates vulnerabilities, offering a novel approach to seamless, secure, and decentralized cross-chain interoperability. A formal security analysis further validates \textit{CrossLink}'s robustness against unauthorized modifications and denial-of-service attacks.
\end{abstract}


\section{Introduction}
As decentralized applications (dApps) expand, there is a growing need for seamless communication between these networks to enable interoperability and more complex functionalities. Cross-chain communication facilitates the exchange of assets, data, and smart contract interactions across multiple blockchain ecosystems, unlocking new possibilities for decentralized finance (DeFi), supply chain management, and multi-chain decentralized applications \cite{ou2022overview}.

Current cross-chain solutions primarily focus on asset transfers \cite{polkadot_whitepaper, cosmos_ibc, layerzero_whitepaper, chainlink_ccip}, enabling users to move tokens between different blockchain networks. These solutions often leverage trusted third-party intermediaries, such as blockchain bridges\cite{lee2023sok} and relay networks\cite{lys2020atomic, westerkamp2022verilay}, to facilitate transactions. While effective for asset bridging, these mechanisms introduce centralization risks, potential points of failure, and security vulnerabilities, as evidenced by several high-profile bridge exploits\cite{lee2023sok}. Despite their utility, these approaches lack a fully decentralized, secure method for executing smart contracts across multiple chains.

Cross-chain smart contract execution is a critical advancement in blockchain interoperability, allowing contracts deployed on different blockchains to communicate and trigger transactions without relying on centralized intermediaries. Achieving this capability would enable decentralized applications to function across multiple networks seamlessly while ensuring consistency, security, and atomicity of transactions\cite{nissl2021towards, westerkamp2022smartsync}. However, designing a trustless, efficient, and secure cross-chain smart contract execution mechanism remains a significant challenge due to issues such as data consistency, state synchronization, and security vulnerabilities.

In this paper, we propose \textbf{\textit{CrossLink}}, a novel framework for decentralized cross-chain smart contract execution. \textit{CrossLink} eliminates the need for trusted third-party intermediaries by leveraging a \textbf{compact chain} that selectively stores curated contract states explicitly authorized for cross-chain interactions. Through a cryptographically secure inter-chain messaging mechanism, \textit{CrossLink} ensures security, correctness and atomic execution while maintaining full decentralization. 

\textit{CrossLink} provides several key benefits over existing cross-chain execution models. It enhances security by implementing a \textbf{deposit/collateral fee mechanism}, mitigating denial-of-service (DoS) attacks and unauthorized data manipulation. Moreover, it ensures data integrity and cross-chain state consistency through efficient state synchronization mechanisms. Our key contributions in this paper include:

\begin{itemize}
    \item We introduce \textit{CrossLink}, a decentralized framework that enables secure cross-chain smart contract execution without reliance on centralized relays or bridges.
    \item We propose a parallel cache blockchain that selectively stores relevant contract states to facilitate secure and controlled cross-chain execution while preserving data confidentiality.
    \item We perform a formal security analysis of \textit{CrossLink}, proving its resilience against unauthorized modifications and DoS attacks.
\end{itemize}

The remainder of this paper is organized as follows: In Section 2, we review related work and identify limitations; in Section 3, we detail the architectural design of CrossLink, including its components and functionality; in Section 4, we discuss implementation details; in Section 5, we evaluate CrossLink’s security and its resilience against attacks; and finally in Section 6, we conclude by summarizing our key contributions and outlining future research directions.
\section{Related Works}
In cross-chain communication research, several approaches have emerged, each with notable limitations. Nissl et al.'s\cite{nissl2021towards} framework for cross-chain smart contract invocation operates exclusively in homogeneous Ethereum environments, significantly restricting its practical application. Similarly, Zala et al.'s\cite{zala2023unlocking} standardized interface approach relies on the assumption that participating blockchains can support standardized interfaces—a requirement that is often infeasible in practice.
IoT-focused research\cite{zhang2023cross} on keyword-based embedded smart contracts neglects to address fundamental scalability concerns critical for implementation. SmartSync's\cite{westerkamp2022smartsync} mirroring approach between master and client contracts depends on identical execution environments, an impractical prerequisite across diverse blockchain ecosystems. Meanwhile, ChainSniper's\cite{tran2024chainsniper} machine learning audit system is hampered by its dependency on training data quality, which varies considerably across different blockchain networks.
Although AtomCI\cite{chen2024atomci} and research on atomic cross-chain execution\cite{yin2025atomic} both attempt to resolve critical consistency issues, they introduce excessive coordination complexity that undermines practical implementation. Despite these advancements, significant challenges remain in achieving fully decentralized, trustless cross-chain smart contract execution, particularly regarding data consistency, state synchronization, and security across heterogeneous blockchain networks.

\section{System Architecture}
\begin{figure}[ht]
  \centering
  \includegraphics[width=3in]{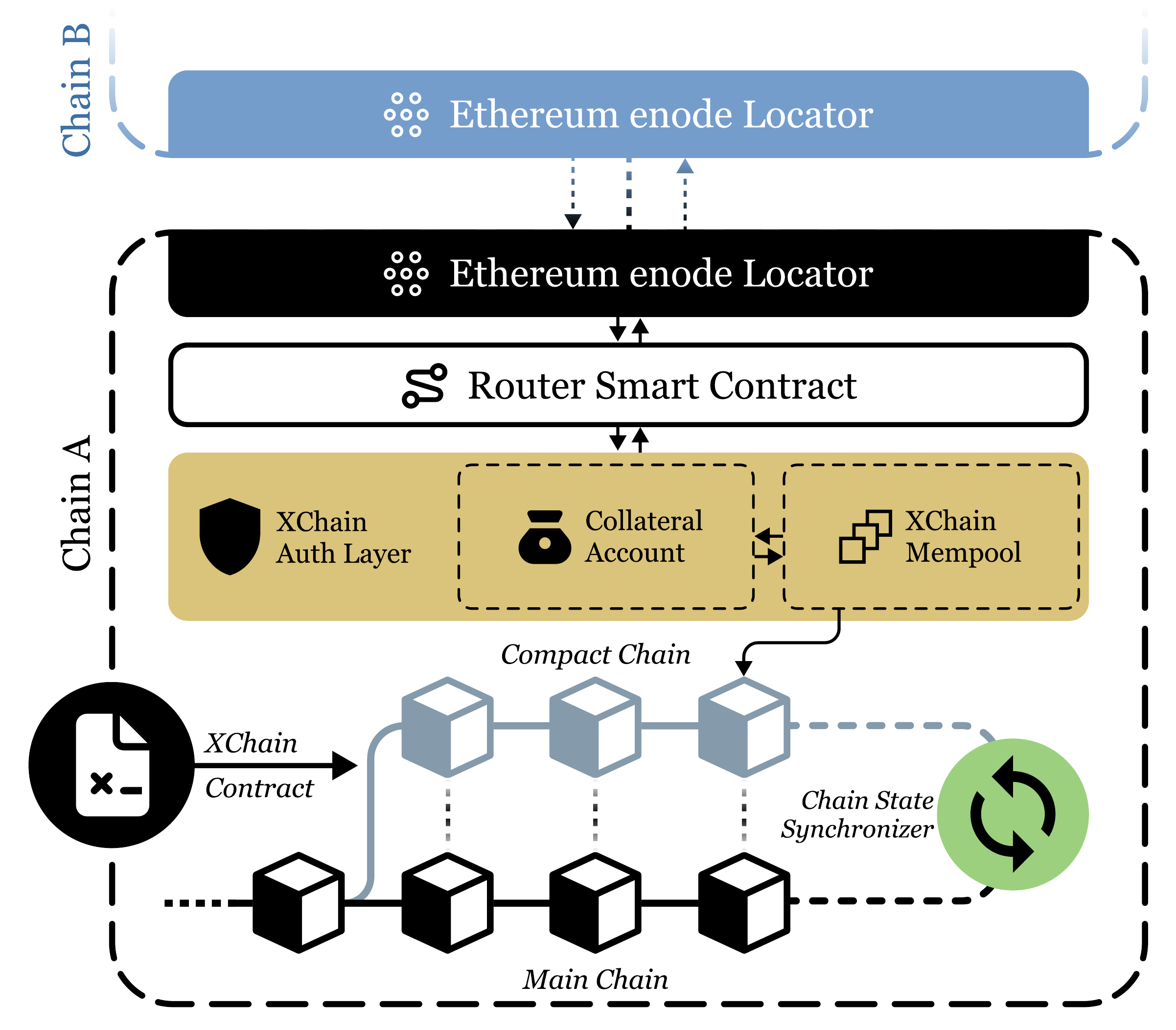}
  \caption{System Architecture of \textit{CrossLink}}
  \label{fig:architecture}
\end{figure}

The system architecture of \textit{CrossLink} is illustrated in Figure~\ref{fig:architecture}. The framework consists of multiple components that enhance cross-chain smart contract invocation while ensuring security and efficiency.

\subsection{Compact Chain Architecture}
The core of \textit{CrossLink}'s design is its Compact Chain, a lightweight sidechain that acts as an intermediary between cross-chain transactions and the Ethereum main chain. Traditional blockchain interoperability mechanisms often struggle with inconsistencies in data structures across networks \cite{zala2023unlocking}, requiring extensive format conversions and validation overhead. To address this, the Compact Chain abstracts publicly accessible data into a standardized, serializable format, enabling efficient synchronization between heterogeneous blockchains.

Instead of executing cross-chain transactions directly on the main chain, which can introduce latency and vulnerability to unforeseen modifications, \textit{CrossLink} processes them first on the Compact Chain. Each block in the Compact Chain stores a reduced form of transaction data, optimized for cross-chain execution, before propagating the results to the main chain. This layered approach improves security, ensures data integrity, and minimizes the risk of state inconsistencies.

\subsection{Chain State Synchronizer}
The Chain State Synchronizer is responsible for maintaining a consistent state between the Ethereum main chain and the Compact Chain. It performs bidirectional state propagation:

\begin{itemize}
    \item When a new block is added to the main chain, the synchronizer extracts public transaction data and generates a corresponding compact block.
    \item When a cross-chain transaction is executed on the Compact Chain, the synchronizer finalizes it by generating an equivalent transaction on the main chain.
\end{itemize}
This dual-layer synchronization strategy ensures that cross-chain operations remain atomic and deterministic, preventing discrepancies between blockchain states. Additionally, the synchronizer implements a transaction validation mechanism that ensures consistency across all interconnected chains.

\subsection{Cross-Chain Authorization Layer}
\textit{CrossLink} employs a dedicated Cross-Chain Authorization (XChain Auth) Layer to ensure that only authenticated and authorized cross-chain transactions are executed. This layer consists of two key components:

\subsubsection{XChain Mempool}
The XChain Mempool acts as an independent transaction pool dedicated to cross-chain transactions. It operates separately from the Ethereum local mempool to prevent conflicts between local and cross-chain transactions. The XChain Mempool follows Ethereum’s standard six-block finalization rule to ensure the security of transactions. However, if a transaction is first added to the Compact Chain and validated, it bypasses the six-block rule, allowing for immediate finalization. This optimization accelerates transaction processing while preserving blockchain security guarantees.

\subsubsection{Collateral Accounts}
To handle gas fees across multiple blockchain networks, \textit{CrossLink} introduces Collateral Accounts. When two blockchains establish a cross-chain communication channel, they each open a collateral account on the counterpart blockchain. This mechanism ensures that users only pay gas fees on their source blockchain, while transaction execution costs on the destination chain are covered by the corresponding collateral account. This approach reduces friction for users and enhances cross-chain transaction efficiency.

\subsection{Router Smart Contract}
The Router Smart Contract is a critical component of the \textit{CrossLink} framework, responsible for facilitating function calls across interconnected blockchains. It is a publicly accessible smart contract deployed on each blockchain network, containing functions that can be invoked by cross-chain smart contracts. The router maintains a list of chain IDs for all interconnected networks and allows users to specify the destination chain and function to execute.

When a user initiates a cross-chain function call, the Router Smart Contract packages the request with the destination chain ID and required parameters, forwarding it to the Ethereum enode locator for transmission. The router ensures seamless function invocation by abstracting the complexities of inter-chain communication, allowing developers to write cross-chain smart contracts without dealing with low-level networking details.

\subsection{Ethereum Enode Locator}
One of the fundamental challenges in decentralized cross-chain communication is identifying and connecting with the correct peer nodes. \textit{CrossLink} leverages Ethereum’s enode system\cite{ethereum_network_addresses} to locate destination chains and establish direct communication channels. The Ethereum enode locator maintains a mapping of enodes (Ethereum node identifiers) to their respective blockchain networks. When a cross-chain request is initiated, the locator resolves the destination chain’s enode and establishes a peer-to-peer HTTP connection to forward the transaction request. By utilizing Ethereum’s existing node discovery mechanisms, \textit{CrossLink} eliminates the need for centralized intermediaries, ensuring fully decentralized cross-chain communication.

\section{Implementation}
We extended the go-ethereum client with custom modules to facilitate cross-chain smart contract invocations\footnote{The source code of CrossLink is available at: \url{https://github.com/CrossLink-Code/CrossLink}}. We modified block processing and synchronization in the Compact Chain architecture to enable bidirectional transaction propagation. For the Cross-Chain Authorization Layer, we added a custom transaction pool and extended account management to support cross-chain collateral lockups. The Router Contract was implemented in Solidity. For the Ethereum enode Locator, we enhanced peer discovery for direct blockchain-to-blockchain communication. We now describe how these components interact to enable cross-chain smart contract execution.

\subsection{Cross-Chain Smart Contract Workflow}
\begin{figure}[ht]
  \centering
  \includegraphics[width=3.2in]{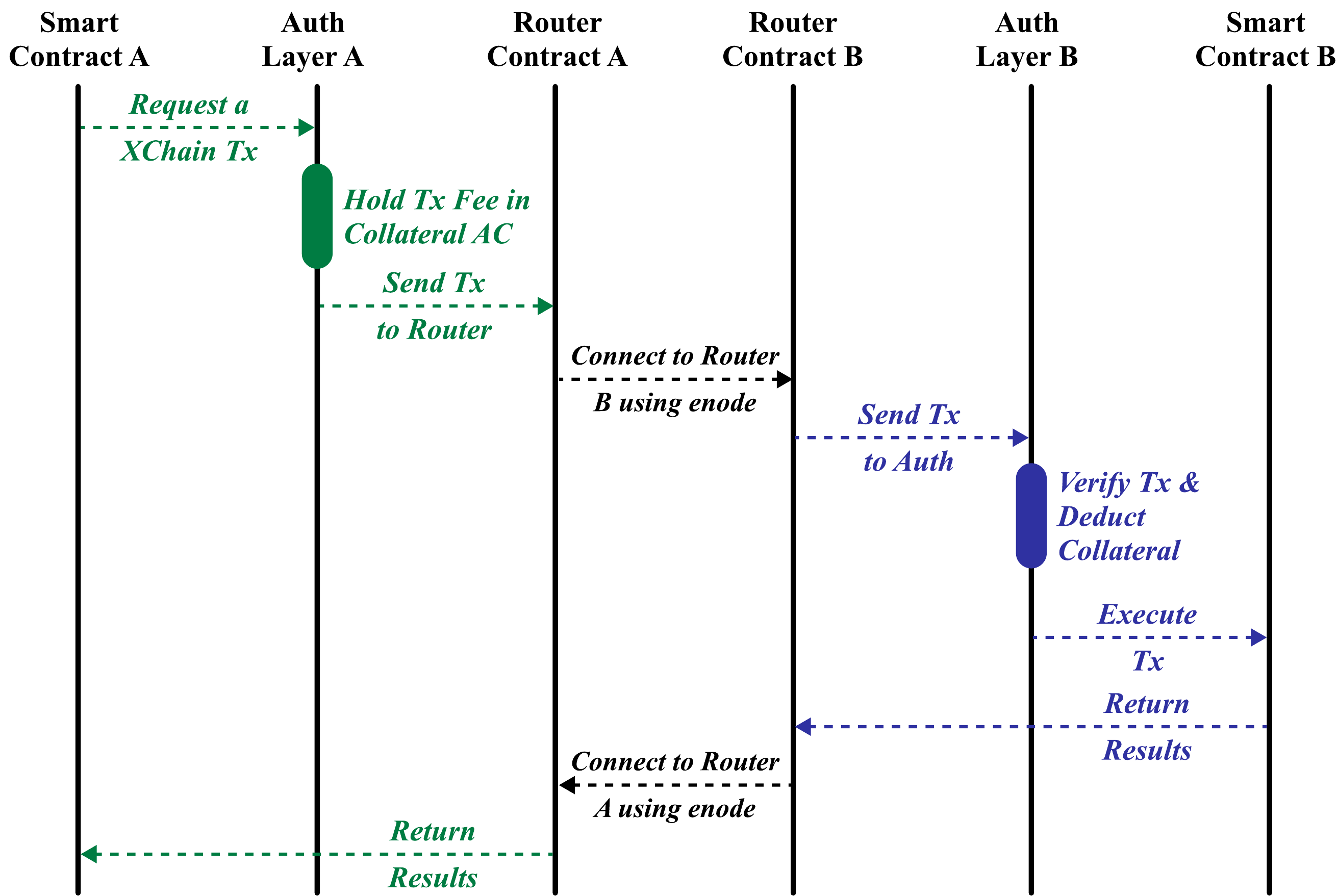}
  \caption{Cross Chain Smart Contract Workflow}
  \label{fig:workflow}
\end{figure}

The workflow of a cross-chain transaction in \textit{CrossLink} is depicted in Figure~\ref{fig:workflow}. The process begins when Smart Contract A on Chain A initiates a cross-chain transaction request by sending it to Auth Layer A. Auth Layer A verifies the request and temporarily holds the transaction fee in a collateral account to ensure that execution costs are covered. Once the transaction is verified, Auth Layer A forwards it to Router Contract A, which is responsible for managing inter-chain communication. Router Contract A utilizes Ethereum's enode mechanism to establish a connection with Router Contract B on Chain B. 

Upon receiving the transaction, Router Contract B forwards it to Auth Layer B for validation. Auth Layer B verifies the transaction details, deducts the required collateral, and ensures that the execution meets the necessary conditions before proceeding. Once validated, Auth Layer B submits the transaction for execution on Smart Contract B, which processes the transaction, updates its state, and generates a result. This result is transmitted back through Auth Layer B. Router Contract B then establishes a reverse enode connection back to Router Contract A. The results pass through Router Contract A and Auth Layer A, where final validation occurs before returning the execution outcome to Smart Contract A. 

This structured execution flow guarantees that cross-chain smart contract interactions are secure, atomic, and efficiently managed. By integrating authentication layers for transaction validation, utilizing router contracts for inter-chain communication, and leveraging enode locators for decentralized peer discovery, \textit{CrossLink} provides a reliable framework for cross-chain interoperability.

\subsection{Cross Chian Interaction Patterns}
\textit{CrossLink} enables reliable smart contract interactions across blockchain networks, supporting both data retrieval and state modifications for seamless interoperability. The following subsections detail fundamental cross-chain operation patterns, demonstrating how smart contracts utilize the framework’s architecture to interact efficiently across multiple blockchains.
\subsubsection{Cross-Chain Data Retrieval Pattern}
In cross-chain data retrieval, Contract A on Blockchain A stores data that Contract B on Blockchain B accesses. The implementation of this process is provided in Appendix B. Contract A defines \texttt{getValue()} to return a stored value, while Contract B implements \texttt{requestValue()} to initiate the cross-chain call and \texttt{handleResult()} to process the response. Contract B calls \texttt{requestValue()}, which interacts with the Router on Blockchain B, invoking its \texttt{initiateCrossChainCall()} function with Contract A’s address, the function selector for \texttt{getValue()}, and its own \texttt{handleResult()} function selector. The Router on Blockchain B emits a \texttt{CrossChainRequest} event, which our modified Ethereum client detects and forwards to Blockchain A. The Router on Blockchain A executes \texttt{getValue()}, retrieves the data, and constructs a callback message, which is then sent back to Blockchain B. Finally, the Router on Blockchain B invokes \texttt{handleResult()} on Contract B, completing the data retrieval process.
\subsubsection{Cross-Chain State Update Pattern }
\textit{CrossLink} enables smart contracts to modify state across blockchain boundaries. The implementation of this process is provided in Appendix C. Contract A defines \texttt{setValue()} for state updates, while Contract B implements \texttt{updateRemoteValue()} to initiate changes and \texttt{handleWriteResult()} for confirmation. Contract B calls \texttt{updateRemoteValue()}, triggering the Router on Blockchain B to execute \texttt{initiateCrossChainCall()} with Contract A’s address, \texttt{setValue()} function selector, and the new value. The Router on Blockchain B emits a \texttt{CrossChainRequest} event, which our modified Ethereum client forwards to Blockchain A. The Router on Blockchain A executes \texttt{setValue()}, updates Contract A’s state, and constructs a callback message. This response is sent back to Blockchain B, where the Router invokes \texttt{handleWriteResult()} on Contract B, confirming the update.

\section{Security Analysis}

Cross-chain smart contract security is critical due to unique challenges absent in single-chain systems. \textit{CrossLink} mitigates these risks by requiring a prepaid execution fee through a collateral smart contract, ensuring pre-coverage of gas costs. This mechanism protects against denial-of-service (DoS) attacks and ensures economic commitment and interoperability.

\textit{\textbf{Lemma-1:}} In \textit{CrossLink}, every cross-chain invocation requires the initiator to lock a prepaid execution fee \( F \) to cover the execution cost \( C_d \) on the destination blockchain. By enforcing a cost function \( f(C_d) \) per invocation, \textit{CrossLink} ensures that large-scale invocation flooding incurs a prohibitive total cost.

\textit{\textbf{Proof:}} Let \( B_s \) and \( B_d \) be the source and destination blockchains, where a cross-chain invocation incurs a computational cost \( c_i \) on \( B_d \). An adversary can flood \( B_d \) with cross-chain invocations \( \{ I_1, I_2, \dots, I_n \} \), creating a total computational load \(\text{Comp}(n) = \sum_{i=1}^{n} c_i\). When \(\text{Comp}(n^*) > \text{Comp}_{\max},\) a \textbf{denial-of-service (DoS) attack} occurs.

\textit{CrossLink} prevents this by enforcing a base fee \( F_{\text{base}} \) via a collateral smart contract, requiring invocations to lock \( F_{\text{base}} \) before execution. The total cost for an attacker executing \( n \) invocations is:
\[
\mathcal{T}(n) = \sum_{i=1}^{n} (F_{\text{base}} + c_i)
\]

Since \( F_{\text{base}} > 0 \), every invocation incurs a minimum cost. For an attacker with capital \( A \), sustaining unlimited invocations requires \( \mathcal{T}(n) \leq A \). However, since costs increase linearly, beyond threshold \( n^* \), the attacker's financial limit is exceeded: \(\mathcal{T}(n^*) > A.\) This establishes an upper bound \( n^* = \max \{ n \mid \mathcal{T}(n) \leq A \} \), making large-scale invocation flooding financially infeasible. Thus, \textit{CrossLink} enforces real economic costs, making DoS attacks impractical.

While economic commitment prevents DoS attacks, securing cross-chain execution also requires protecting contract state integrity. A key risk is malicious actors exploiting cross-chain interactions to manipulate contract data, leading to inconsistencies or unauthorized modifications. \textit{CrossLink} mitigates this by using the Compact Chain to separate locally modifiable data from cross-chain accessible data, ensuring secure interactions while preventing unauthorized changes 

\textit{\textbf{Lemma-2:}} The Compact Chain \( B_{\text{compact}} \) contains only \( S_{\text{compact}} \), which stores data specifically for cross-chain interactions. The Main Chain \( B_{\text{main}} \) maintains both \( S_{\text{main}} \), which holds contract state for internal execution, and \( S_{\text{compact}} \), ensuring synchronization between the two chains. Since cross-chain transactions \( X \) modify only \( S_{\text{compact}} \), external actors invoking cross-chain transactions cannot manipulate \( S_{\text{main}} \), ensuring data integrity and preventing malicious cross-chain modifications.

\textit{\textbf{Proof:}} The Main Chain $B_{\text{main}}$ maintains the full contract state, consisting of $S_{\text{main}}$ which is the internal execution state and $S_{\text{compact}}$ which is the cross-chain interaction state. The Compact Chain $B_{\text{compact}}$ maintains only $S_{\text{compact}}$, ensuring that cross-chain transactions operate within an isolated storage space. For any cross-chain transaction $X$ executed on $B_{\text{compact}}$, the state transition is restricted to:
\( X: S_{\text{compact}} \rightarrow S_{\text{compact}}' \)

Since $S_{\text{main}}$ exists only on $B_{\text{main}}$ and cross-chain transactions are executed exclusively on $B_{\text{compact}}$, access to $S_{\text{main}}$ is formally restricted: \( \text{Access}(X) \cap S_{\text{main}} = \emptyset \). This isolation ensures that for any cross-chain transaction $X$, it holds that
\( X: S_{\text{main}} \rightarrow S_{\text{main}} \). Since $S_{\text{main}}$ remains unchanged under all valid executions of $X$, no valid execution path exists for an external actor to modify $S_{\text{main}}$, ensuring data integrity and preventing malicious cross-chain modifications.

\section{Conclusion}
In this paper, we introduced a decentralized framework facilitating trustless cross-chain smart contract execution, \textit{CrossLink}. It addresses the security and centralization concerns of current solutions. By utilizing selective state storage through a compact chain and secure messaging, along with a deposit/collateral system, \textit{CrossLink} ensures resilience and data integrity. Our security analysis validates its potential to revolutionize interoperability. Looking ahead, we aim to extend \textit{CrossLink}'s implementation beyond the Ethereum ecosystem, exploring its compatibility and performance across diverse blockchain architectures. This expansion will involve rigorous real-life performance evaluations to assess its scalability and efficiency in various network conditions. Furthermore, we plan to develop and integrate specific functionalities that empower developers to build sophisticated cross-chain dApps, enabling seamless interaction and data exchange across multiple blockchain environments. This will unlock new possibilities for decentralized applications, fostering a more interconnected and efficient blockchain ecosystem.

\bibliographystyle{IEEEtran}
\bibliography{IEEEabrv,references}
\appendices

\section{Cross-Chain Router Implementation}
\small This code is also available at: \textcolor{blue}{\url{https://github.com/CrossLink-Code/CrossLink/blob/main/CrossChainRouter.sol}}
\begin{lstlisting}[
    language=Solidity, 
    caption=Cross-Chain Router Contract, 
    label=lst:crosschainrouter,
    numbers=left,
    numberstyle=\tiny\color{black},
    basicstyle=\scriptsize\ttfamily,
    breaklines=true,
    xleftmargin=2em, 
    framexleftmargin=2em,
    xrightmargin=1em, 
    framexrightmargin=1em,
    tabsize=2
]
pragma solidity ^0.8.19;

import "./ICrossChainRouter.sol";

contract CrossChainRouter is ICrossChainRouter {
    mapping(bytes32 => Blockchain) public blockchains;
    bytes32 public immutable routerChainId;

    constructor(bytes32 _chainId) {
        routerChainId = _chainId;
    }

    function chainId() external view override returns (bytes32) {
        return routerChainId;
    }

    function initiateCrossChainCall(
        bytes32 _targetChainId,
        ExternalContract calldata _targetContract,
        Callback calldata _callback
    ) external payable override {
        bytes32 requestId = keccak256(abi.encodePacked(block.timestamp, msg.sender));

        emit CrossChainRequest(
            requestId,
            blockchains[_targetChainId],
            CrossChainCall({
                requestId: requestId,
                sender: msg.sender,
                targetContract: _targetContract,
                callback: _callback
            })
        );
    }

    function handleIncoming(CrossChainCall calldata crossChainCall) external override {
        require(crossChainCall.targetContract.contractAddress != address(0), "Invalid contract address");

        (bool success, bytes memory returnData) = crossChainCall.targetContract.contractAddress.call(
            abi.encodePacked(crossChainCall.targetContract.functionSelector, crossChainCall.targetContract.params)
        );

        if (success) {
            if (crossChainCall.callback.callbackAddress == address(0)) {
                return;
            }

            if (crossChainCall.callback.chain.chainId == routerChainId) {
                (bool status, bytes memory result) = crossChainCall.callback.callbackAddress.call(
                    abi.encodePacked(crossChainCall.callback.callbackSelector, returnData)
                );
                emit CallBackResult(crossChainCall.requestId, status, result);
            } else {
                ExternalContract memory extContract = ExternalContract({
                    contractAddress: crossChainCall.callback.callbackAddress,
                    functionSelector: crossChainCall.callback.callbackSelector,
                    params: returnData
                });

                this.initiateCrossChainCall(
                    crossChainCall.callback.chain.chainId,
                    extContract,
                    Callback({ chain: Blockchain(bytes32(0), ""), callbackAddress: address(0), callbackSelector: bytes4(0) })
                );
            }
        }
    }
}
\end{lstlisting}

\section{Read Example}

\subsection{Contract on Chain A (Provides Data)}
\small This code is also available at: \textcolor{blue}{\url{https://github.com/CrossLink-Code/CrossLink/blob/main/ContractA_Read.sol}}
\begin{lstlisting}[
    language=Solidity, 
    caption=ContractA\_Read: Provides Data, 
    label=lst:contractA_read,
    numbers=left,
    numberstyle=\tiny\color{black},
    basicstyle=\scriptsize\ttfamily,
    breaklines=true,
    xleftmargin=2em, 
    framexleftmargin=2em,
    xrightmargin=1em, 
    framexrightmargin=1em
]
// SPDX-License-Identifier: MIT
pragma solidity ^0.8.19;

contract ContractA_Read {
    uint256 public storedValue = 42;

    function getValue() external view returns (uint256) {
        return storedValue;
    }
}
\end{lstlisting}

\subsection{Contract on Chain B (Reads Data)}
\small This code is also available at: \textcolor{blue}{\url{https://github.com/CrossLink-Code/CrossLink/blob/main/ContractB_Read.sol}}
\begin{lstlisting}[
    language=Solidity, 
    caption=ContractB\_Read: Reads Data via CrossChainRouter, 
    label=lst:contractB_read,
    numbers=left,
    numberstyle=\tiny\color{black},
    basicstyle=\scriptsize\ttfamily,
    breaklines=true,
    xleftmargin=2em, 
    framexleftmargin=2em,
    xrightmargin=1em, 
    framexrightmargin=1em
]

pragma solidity ^0.8.19;

import "./ICrossChainRouter.sol";

contract ContractB_Read {
    ICrossChainRouter public router;
    uint256 public retrievedValue;

    constructor(address _router) {
        router = ICrossChainRouter(_router);
    }

    function requestValue(bytes32 targetChainId, address targetContract) public {
        ICrossChainRouter.ExternalContract memory target = ICrossChainRouter.ExternalContract({
            contractAddress: targetContract,
            functionSelector: bytes4(keccak256("getValue()")),
            params: ""
        });

        ICrossChainRouter.Callback memory callback = ICrossChainRouter.Callback({
            chain: ICrossChainRouter.Blockchain(targetChainId, ""),
            callbackAddress: address(this),
            callbackSelector: bytes4(keccak256("handleResult(uint256)"))
        });

        router.initiateCrossChainCall(targetChainId, target, callback);
    }

    function handleResult(uint256 value) external {
        retrievedValue = value;
    }
}
\end{lstlisting}

\section{Write Example}

\subsection{Contract on Chain A (Allows Updates)}
\small This code is also available at: \textcolor{blue}{\url{https://github.com/CrossLink-Code/CrossLink/blob/main/ContractA_Write.sol}}
\begin{lstlisting}[
    language=Solidity, 
    caption=ContractA\_Write: Allows Data Updates, 
    label=lst:contractA_write,
    numbers=left,
    numberstyle=\tiny\color{black},
    basicstyle=\scriptsize\ttfamily,
    breaklines=true,
    xleftmargin=2em, 
    framexleftmargin=2em,
    xrightmargin=1em, 
    framexrightmargin=1em
]
// SPDX-License-Identifier: MIT
pragma solidity ^0.8.19;

contract ContractA_Write {
    uint256 public storedValue = 0;

    function setValue(uint256 _newValue) external returns (bool) {
        storedValue = _newValue;
        return true;
    }
}
\end{lstlisting}

\subsection{Contract on Chain B (Modifies Data on Chain A)}
\small This code is also available at: \textcolor{blue}{\url{https://github.com/CrossLink-Code/CrossLink/blob/main/ContractB_Write.sol}}
\begin{lstlisting}[
    language=Solidity, 
    caption=ContractB\_Write: Modifies Data via CrossChainRouter, 
    label=lst:contractB_write,
    numbers=left,
    numberstyle=\tiny\color{black},
    basicstyle=\scriptsize\ttfamily,
    breaklines=true,
    xleftmargin=2em, 
    framexleftmargin=2em,
    xrightmargin=1em, 
    framexrightmargin=1em
]
// SPDX-License-Identifier: MIT
pragma solidity ^0.8.19;

import "./ICrossChainRouter.sol";

contract ContractB_Write {
    ICrossChainRouter public router;
    bool public writeSuccessful;

    constructor(address _router) {
        router = ICrossChainRouter(_router);
    }

    function updateRemoteValue(bytes32 targetChainId, address targetContract, uint256 newValue) public {
        ICrossChainRouter.ExternalContract memory target = ICrossChainRouter.ExternalContract({
            contractAddress: targetContract,
            functionSelector: bytes4(keccak256("setValue(uint256)")),
            params: abi.encode(newValue)
        });

        ICrossChainRouter.Callback memory callback = ICrossChainRouter.Callback({
            chain: ICrossChainRouter.Blockchain(targetChainId, ""),
            callbackAddress: address(this),
            callbackSelector: bytes4(keccak256("handleWriteResult(bool)"))
        });

        router.initiateCrossChainCall(targetChainId, target, callback);
    }

    function handleWriteResult(bool success) external {
        writeSuccessful = success;
    }
}
\end{lstlisting}

\end{document}